# Reinforcement Learning for Process Control with Application in Semiconductor Manufacturing

Yanrong Li[1], Juan Du[2] and Wei Jiang[1]

[1]Antai College of Economics and Management, Shanghai Jiao Tong University, Shanghai, China

[2]Smart Manufacturing Thrust, Systems Hub, The Hong Kong University of Science and Technology, Guangzhou, China

**Abstract**

Process control is widely discussed in the manufacturing process, especially for semiconductor manufacturing. Due to unavoidable disturbances in manufacturing, different process controllers are proposed to realize variation reduction. Since reinforcement learning (RL) has shown great advantages in learning actions from interactions with a dynamic system, we introduce RL methods for process control and propose a new controller called RL-based controller. Considering the fact that most existing process controllers mainly rely on a linear model assumption for the process input–output relationship, we first propose theoretical properties of RL-based controllers according to the linear model assumption. Then the performance of RL-based controllers and traditional process controllers (e.g., exponentially weighted moving average (EWMA), general harmonic rule (GHR) controllers) are compared for linear processes. Furthermore, we find that the RL-based controllers have potential advantages to deal with other complicated nonlinear processes that are with and without assumed explicit model formulations. The intensive numerical studies validate the advantages of the proposed RL-based controllers.

## 1. Introduction

Process control is necessary for reducing the variations in manufacturing processes to improve the quality and productivity of the final products. For example, in the semiconductor manufacturing process, different unavoidable disturbances (e.g., tool-induced and product-induced disturbances) caused by various manufacturing environments could influence the stability of the manufacturing process and the quality of the final products (Su et al., 2007). Therefore, designing an efficient control strategy that considers various disturbances to reduce variation is a significant research problem in the semiconductor manufacturing process.

In practice, sensors installed in production lines can collect various data during the manufacturing process including the system's input control actions and output quality data. Then the control recipes can be designed and optimized to compensate for disturbances (Castillo and Hurwitz, 1997). In the semiconductor manufacturing process, existing research can be categorized into two groups from theoretical and practical perspectives of controller designs to realize variation reduction and quality improvement.

In terms of the theoretical design of controllers, extensive research generally made assumptions on the process model with *predefined* disturbances followed by corresponding designs of the controller for the semiconductor manufacturing process. For example, Sachs et al. (1995) introduced the Exponentially Weighted Moving Average (EWMA) controller for the integrated moving average (IMA) disturbance process. Based on this work, more complicated disturbance processes such as Autoregressive Moving Average (ARMA) and Autoregressive Integrated Moving Average (ARIMA) processes are discussed. For example, Tsung and Shi (1999) focused on the ARMA (1,1) disturbance and proposed Proportional-Integral-Derivative (PID) controller to deal with ARMA disturbance. Tseng et al. (2003) proposed the variable-EWMA (VEWMA) controller to optimize the discount factor of EWMA in the ARIMA process and found that the VEWMA controller is easy to implement by calculating the optimal discount factor. Furthermore, more general controllers are also proposed to deal with various kinds of disturbances. For example, He et al. (2009) proposed a General Harmonic Rule (GHR) controller and proved that the GHR controller could handle IMA(1,1), ARMA (1,1), and ARIMA (1,1) disturbances quite well. These theoretical controllers provide a foundation for process control in semiconductor manufacturing, which inspires many extensions in industry practice.

In terms of practical extensions, specific applications are considered in the control scheme according to the manufacturing scenarios. For example, Wang and Han (2013) proposed a batch-based EWMA controller by considering the batch production of semiconductors manufacturing. Liu et al. (2018) reviewed comprehensive literature of the batch-based process and summarized control principles and simulation examples of various controllers. Huang and Lv (2020) improved the EWMA controller by considering the online measurement in the batch production process. In addition to the considerations of batch-based characteristics, other cases are also discussed. For example, Djurdjanović et al. (2017) proposed a robust automatic control method based on inaccurate knowledge about process noises and applied it in the lithography processes.

Existing works make influential contributions to semiconductor manufacturing, and a linear process model with certain disturbance is usually assumed. In practice, the process model may not be linear, and the disturbance can be different from the assumption. Therefore, using a *predefined* fixed linear model with one type of disturbance to describe the input–output relationship can experience difficulties in fitting the entire manufacturing procedure accurately (Wang and Shi, 2020). A more flexible controller that can be applied in various cases is desired.

Reinforcement learning (RL) is a powerful data-driven method to learn actions in dynamic environments or systems without assuming process models or disturbances (Kaelbling et al., 1996). RL decides to minimize the total cost by learning the relationship between the input actions and outputs directly from historical actions and output data (Sutton and Barto, 1998). Due to the comprehensive consideration of the real-time system output and historical control strategies, RL-based control methods have the potential to handle various cases in manufacturing process control problems. In addition, RL has already shown a great advantage in robotics control problems (Kober, 2013). However, to our best knowledge, few works deal with process control problems in semiconductor manufacturing by RL. Therefore, we fill the research gap by developing RL-based controllers for process control in semiconductor manufacturing.

Our contributions can be summarized as follows: (1) we first proposed RL-based controllers, which can be applied in different cases. If the domain knowledge is available, RL-based controllers with different approximate process models (such as linear or nonlinear) are proposed. Otherwise, if there is no evidence to assume a proper model, the RL-based controller with policy gradient search method can be applied; (2) Two computational algorithms for RL-based controllers are presented according to whether domain knowledge is available or not; (3) Theoretical properties are investigated for RL-based controllers given the assumption of the widely accepted linear process models.

The remainder of this paper is organized as follows. Section 2 provides the methodology of RL-based controllers including the model formulations, algorithms, and theoretical properties. For fair comparisons with conventional controllers, Section 3 introduces two classical linear simulation case studies in the semiconductor manufacturing process. To further validate the performance of RL-based controllers, Section 4 presents two simulation cases with other complicated nonlinear process models, where traditional controllers are inapplicable. Section 5 summarizes the conclusions and future research.

## 2. Methodology of RL in process control

In this section, we first introduce the formulation of the process control problems. Then, the methodology of RL-based controllers is developed. Specifically, if the domain knowledge is available for process model approximation, the RL-based controller with approximate models is proposed in Algorithm 1. Otherwise, we propose the RL-based controller with policy gradient search in Algorithm 2 to find the optimal control strategies. In addition, theoretical properties are analyzed to guarantee the performance of the RL-based controller under the generalized linear assumption.

### 2.1 *Problem definition and formulation*

We first denote the set of all sampling time points of the system output as $T$, and $T$ is a finite set (i.e., $T = \{0,1,2,\ldots t,\ldots,T\}$). Then, the total manufacturing process within a run is divided into $T$ periods, where the system output can be observed during each period. The data collected from time 0 to $T$ is defined as a sample path. Denote $y_t$ as the system output at time $t$, and the initial system output as $y_0$. Following the existing process control formulations (Del Castillo and Hurwitz, 1997; He et al. 2009; Wang and Han, 2013), we define the process model as $y_t = f_t(y_{t-1}, u_t, d_t)$ for $t \in T$, where $d_t$ is the process disturbance of the system at time $t$, and $u_t$ is the control action at period $t$. Figure 1 illustrates process control in the semiconductor manufacturing process. In each period, there is an unavoidable disturbance $d_t$, which can influence the system output, and the control action $u_t$ aims to compensate for the effect of $d_t$.

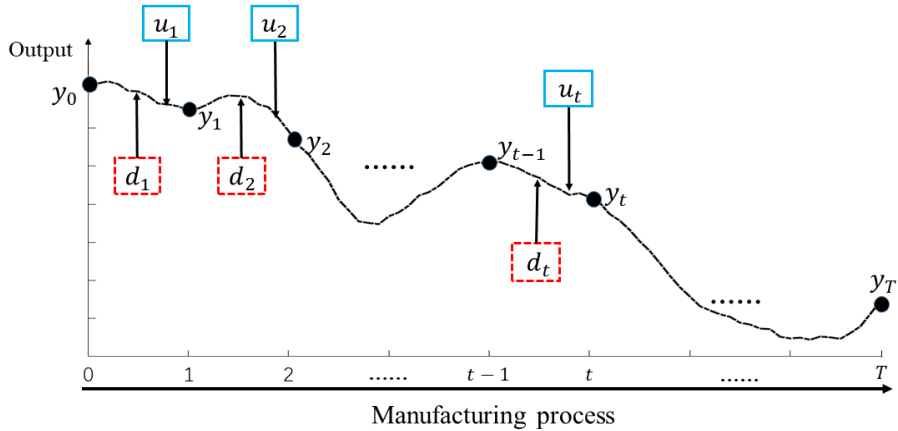

Figure 1. Illustration of semiconductor manufacturing control.

Our objective is to keep $y_t$ close to the target level $y^*$ in each period $t \in \{1,2,\ldots,T\}$ in terms of squares loss to achieve a higher level of process capability and quality (Wang and Han, 2013). In summary, we would like to obtain the control action $u_t$ by solving the optimization problem as follows:

$$\min_{u_t} \sum_{t=1}^{T}(y_t - y^*)^2$$

$$\text{s.t. } y_t = f_t(y_{t-1}, u_t, d_t), \quad (2.1)$$

where $d_t$ is the disturbance at period $t$, which is a random variable related to the historical disturbances and manufacturing periods. $f_t(\cdot)$ is the process model, which is often assumed by a linear model with a specific series of disturbances in the literature of semiconductor manufacturing process control. Generally, if the process model is unknown, the underlying model needs to be discussed. One of the most effective solutions is to find an approximate process model $\hat{f}_t(\cdot)$, and minimize the difference between $f_t(\cdot)$ and $\hat{f}_t(\cdot)$. Since $f_t(\cdot)$ is unknown, we use $y_t$ as the observation of $f_t(\cdot)$ in practice, i.e., the optimization problem is:

$$\min_{\hat{f}_t} \sum_{t=1}^{T}\|y_t - \hat{f}_t(\cdot)\|^2. \quad (2.2)$$

Theoretically, the approximate function $\hat{f}_t(\cdot)$ can be any form and a proper $\hat{f}_t(\cdot)$ is necessary to describe the real system but difficult to be determined. Therefore, in the following subsections, we discuss two scenarios to specify $\hat{f}_t(\cdot)$ with and without domain knowledge, respectively.

## 2.2 RL-based controller with domain knowledge

In this section, we discuss RL-based controllers given domain knowledge. First, the general case of the process model is investigated, and the corresponding algorithm is proposed. Second, several theorems are proposed given the most widely accepted linear process model. Finally, a special case of the time-independent process model is analyzed.

If the domain knowledge is available for us to estimate $f_t(\cdot)$ by a specific model $\hat{f}_t(\cdot|\boldsymbol{\Theta}_t)$ with a parameter set denoted as $\boldsymbol{\Theta}_t$, Algorithm 1 considers the general case that the process model is dynamic in different periods. After determining the formulation of the approximated function $\hat{f}_t(\cdot|\boldsymbol{\Theta}_t)$, RL-based controller estimates the parameters of the model according to different sample paths that are collected and stored in the historical offline dataset $\boldsymbol{D}$. Statistical inference methods such as maximum likelihood estimation can be used. Then, according to the parameter estimators in $\hat{f}_t(\cdot|\widehat{\boldsymbol{\Theta}}_t)$, the optimal

control action $\hat{u}_t^*$ can be determined. After executing the action $\hat{u}_t^*$ and observing the output $\hat{y}_t^*$, both action $\hat{u}_t^*$ and output $\hat{y}_t^*$ are collected and added into the offline dataset $\boldsymbol{D}$ to re-estimate the parameters. Algorithm 1 repeats the procedures of optimizing control actions and estimating parameters alternately until the convergence of the optimal action and parameters in the approximate function (i.e. the difference between adjacent iterated values of $\hat{u}_t$ and $\widehat{\boldsymbol{\Theta}}_t$ is smaller than the corresponding thresholds $\varepsilon$ and $\eta$).

---

Algorithm 1. RL-based controller with domain knowledge

---

Input: function $\hat{f}_t(\cdot|\boldsymbol{\Theta}_t)$, parameter $T$, $y^*$, $\boldsymbol{\varepsilon}$, $\eta$

Initialize $y_0$, $T$, $u_1^0$, $\boldsymbol{D} = \{D_1, D_2, \dots D_T\}$

Execute the initial control strategy $u_1^0$ and record the outputs $y_1^0$

**for** $t = 1:T$ **do**

    add $(y_{t-1}^0,\ u_t^0,\ y_t^0)$ to dataset $D_t$

    **Repeat**:

        $\min_{\boldsymbol{\Theta}_t^k} \sum_{i=0}^{k}\left\|\hat{f}_t(y_{t-1}^i, u_t^i; \boldsymbol{\Theta}_t^k) - y_t^i\right\|^2$ based on dataset $D_t$

        $\min_{u_t^{k+1}}\left(\hat{f}_t(y_{t-1}^{k+1}, u_t^{k+1}; \boldsymbol{\Theta}_t^k) - y^*\right)^2$

        execute $u_t^{k+1}$ and record the outputs $y_t^{k+1}$

        add $(y_{t-1}^{k+1},\ u_t^{k+1},\ y_t^{k+1})$ to dataset $D_t$

    **Until** $\left\|\boldsymbol{\Theta}_t^{k+1} - \boldsymbol{\Theta}_t^k\right\| < \varepsilon$, and $\left\|u_t^{k+1} - u_t^k\right\| < \eta$

    Record the final action $u_t$ and $y_t$

    $u_{t+1}^0 \leftarrow u_t$ if $t \neq T$

**end for**

---

Since there are many choices for the form of $\hat{f}_t(\cdot)$, and it is infeasible to analyze all the possible models. Here, we especially analyze the most widely accepted generalized linear model assumptions on parameters (Theorem 1), and control actions (Theorem 2) in the semiconductor manufacturing process. The details are as follows. All the proofs are listed in the Appendix.

**Assumption 2.1.** *Suppose that $f_t(\cdot)$ is well-approximated by $\hat{f}_t(\cdot|\boldsymbol{\Theta}_t)$, which can be formulated as* $\hat{f}_t(\cdot|\boldsymbol{\Theta}_t) = \boldsymbol{\Theta}_t \boldsymbol{\Psi}_t(\boldsymbol{\tau}_t) + e_t$, where $\boldsymbol{\tau}_t = [u_t, u_{t-1} \ldots u_1, d_{t-1} \ldots d_1, 1]$ is the trajectory of control actions and disturbances, $\boldsymbol{\Psi}_t(\cdot)$ is an arbitrary function of $\boldsymbol{\tau}_t$, and $e_t$ is defined as independent normally distributed errors with zero mean.

**Theorem 1.** *Based on Assumptions.2.1, let $\widehat{\boldsymbol{\Theta}}_t$ denote the parameter estimators in $\hat{f}_t(\cdot|\boldsymbol{\Theta}_t)$, we have: $E(\widehat{\boldsymbol{\Theta}}_t - \boldsymbol{\Theta}_t) = \mathbf{0}$ and $\mathrm{var}(\widehat{\boldsymbol{\Theta}}_t - \boldsymbol{\Theta}_t) \sim O(1/N)$, where N is the total number of sample paths.* The proof is provided in Appendix A.

Theorem 1 guarantees the convergence order of the RL-based controller in Algorithm 1 based on the dynamic process model specified by the generalized linear model. Different sample paths are considered to estimate parameters in $\hat{f}_t(\cdot|\boldsymbol{\Theta}_t)$, and make control decisions according to $\hat{f}_t(\cdot|\widehat{\boldsymbol{\Theta}}_t)$. According to Theorem 1, we have the variance of the difference between parameter estimators $\widehat{\boldsymbol{\Theta}}_t$ and their ground truth $\boldsymbol{\Theta}_t^*$ is derived to be the order of $O(1/N)$. If $f_t(\cdot)$ can be well-approximated by $\hat{f}_t(\cdot|\boldsymbol{\Theta}_t)$, with the converged $\widehat{\boldsymbol{\Theta}}_t$, control action optimization based on $\hat{f}_t(\cdot|\widehat{\boldsymbol{\Theta}}_t)$ will also converge to that based on $f_t(\cdot|\boldsymbol{\Theta}_t^*)$, which is the real optimal control action. Specifically, if $\boldsymbol{\Psi}_t(\cdot)$ is also a linear function with current control action $u_t$, more theoretical properties can be analyzed. Therefore, we have the further assumption and theorem as follows.

**Assumption 2.2.** *The approximate process model can be specified as a linear function with current control actions, which is formulated as $\hat{f}_t(\cdot|\boldsymbol{\Theta}_t) = b_t u_t + \boldsymbol{Y}_t \boldsymbol{K}_t\left(\boldsymbol{\tau}_t^{[\backslash u_t]}\right) + e_t$, where $\boldsymbol{\Theta}_t^T = [b_t, \boldsymbol{Y}_t^T]$, $\boldsymbol{\tau}_t^{[\backslash u_t]} = [u_{t-1} \ldots u_1, d_{t-1} \ldots d_1, 1]$ denotes the trajectory of control actions and disturbances except $u_t$, $\boldsymbol{K}_t(\cdot)$ is an arbitrary function of $\boldsymbol{\tau}_t^{[\backslash u_t]}$, and $e_t$ is defined as independent normally distributed errors with zero mean.*

By reformulating the approximate process model in Assumption 2.2, we have the expression of $\hat{f}_t(\cdot|\boldsymbol{\Theta}_t)$ as follows:

$$y_t = \hat{f}_t(\cdot|\boldsymbol{\Theta}_t) \begin{aligned} &= b_t u_t + \boldsymbol{Y}_t \boldsymbol{K}_t\left(\boldsymbol{\tau}_t^{[\backslash u_t]}\right) + e_t \\ &= b_t u_t + c_t + e_t \end{aligned} \quad (2.3)$$

where $c_t = \boldsymbol{Y}_t \boldsymbol{K}_t\left(\boldsymbol{\tau}_t^{[\backslash u_t]}\right)$. The parameter estimators of $c_t$ and $b_t$ are denoted by $\hat{c}_t$ and $\hat{b}_t$ respectively. Then, we will have the following Theorem 2.

**Theorem 2.** *In each period $t$, let $\mu_1$ and $\mu_2$, $\sigma_1$ and $\sigma_2$ denote the mean values and standard deviations of $(y^* - \hat{c}_t)$ and $\hat{b}_t$ respectively, and $\sigma_{12}$ denotes the covariance of $(y^* - \hat{c}_t)$ and $\hat{b}_t$. We have the upper bound on the probability of the control errors of the weighted sample mean $\bar{u}_t = K_t\left(\tau_t^{[\backslash u_t]}\right)\left(K_t^T K_t\right)^{-1} K_t^T U_t$ and its corresponding output $y_t$ as follows:*

$$\begin{cases} P(|\bar{u}_t - u_t^*| > \eta) \leq \frac{\sigma_2^2 \sigma_1^2 - \sigma_{12}^2}{(\mu_2 \sigma_2 \eta)^2} + 2\Phi\left(-\frac{\mu_2}{\sigma_2}\right) \\ P\left(|E(y_t(\bar{u}_t)) - y_t^*| > \eta\right) \leq \frac{\sigma_2^2 \sigma_1^2 - \sigma_{12}^2}{(\sigma_2 \eta)^2} + 2\Phi\left(-\frac{\mu_2}{\sigma_2}\right), \end{cases} \quad (2.4)$$

*where $\Phi(\cdot)$ is the cumulative distribution function of standard normal distribution; $K_t\left(\tau_t^{[\backslash u_t]}\right)$ is the function on the online historical trajectory of control actions and disturbances before period $t$; $U_t$ is the offline control action at period $t$ and $K_t$ is the observation of the function on the offline historical trajectory actions and disturbances before period $t$.*

The proof is provided in Appendix B.

With the increased number of sample paths, the variance of parameters will be reduced, the ratio between variance and mean will be reduced. Therefore, the upper bound of the difference between the estimated control action (output) and the optimal control action (output) will also be reduced. Theorem 2 proposes theoretical error bounds of control actions and corresponding control costs for the widely used linear process model in semiconductor manufacturing.

In many practical applications in process control problems, the process model is assumed to be time-independent, i.e., the transition function $\hat{f}_t(\cdot | \boldsymbol{\Theta}_t)$ can be simplified by $\hat{f}(\cdot | \boldsymbol{\Theta}_t)$. The output $y_t$ and action $u_t$ at any period $t$ ($t \in \{1,2,\dots,T\}$) can be used to estimate $\hat{f}(\cdot | \boldsymbol{\Theta}_t)$. Therefore, even in a single sample path, with the increase of $T$, the parameter estimators converge to their real values.

Notably, traditional statistical methods tend to estimate the parameters according to samples at first, and then optimize the control action. Compared with the traditional method, RL-based control has a unique characteristic: "learning-by-doing". In each iteration, according to the estimated parameters in the approximate function $\hat{f}(\cdot | \boldsymbol{\Theta}_t)$, the control decision is optimized. Then, the current decision and the output observation are used to re-estimate the parameters. Therefore, the current optimal control action contributes to reinforcing the knowledge on decision making.

Specifically, if $f_t(\cdot)$ is time-independent, following Assumptions 2.1 and 2.2, the output of RL-

based controller in Algorithm 1 has less variation than the method of making optimization after parameter estimation. For example, if $y_t = a + bu_t + e_t$, even in a single sample path, we have the variance of output prediction $\hat{y}_t$ given a certain control action $u_t$ as:

$$\text{var}(\hat{y}_t) = \left( \frac{1}{t-1} + \frac{(u_t - \bar{u})^2}{\sum_{i=1}^{t-1}(u_i - \bar{u})^2} \right) \sigma^2, \tag{2.5}$$

where $\bar{u}$ is the mean of control actions until period $t-1$, and $\sigma^2$ is the variance of the random error $e_t$. When $f_t(\cdot)$ is time-independent, which can be denoted as $f(\cdot)$, with the increase of the sample length from period 1 to $t-1$, parameter estimators will converge to the real values and the control action $u_t$ also converges to the real optimal control action $u^*$ as it is optimized according to parameter estimators. Then the mean of historical control actions $\bar{u}$ also converges to $u^*$. Therefore, comparing with random sampling in the traditional optimization after parameter estimation method, "learning-by-doing" in RL-based controller can reduce the variance of $\hat{y}_t$ by reducing the difference of $u_t$ and $\bar{u}$ in Equation (2.5). This property makes it easy for RL-based controllers to find the optimal control action based on $\hat{y}_t$, especially when samples are limited.

In summary, Theorem 1 guarantees the convergency rate of parameters in the generalized linear process models for RL-based controllers. Theorem 2 proposes the upper bound on the probability of the control errors. In addition, an important advantage of RL-based controllers based on the time-independent process model is discussed. Furthermore, if the process model is nonlinear, Recht (2019) discussed that as long as $\hat{f}_t(\cdot)$ is consistent with the real function $f_t(\cdot)$, the RL-based controller with domain knowledge has good performance. While if the wrong form of $\hat{f}_t(\cdot)$ is chosen, then the inconsistent form will lead to the bias of parameter estimation and then increases the variation of system output in the control process. Therefore, if a reasonable or accurate approximate process model is unavailable, we propose the RL-based controller without domain knowledge, as shown in Section 2.3.

## 2.3 RL-based controller without domain knowledge

To deal with the cases that domain knowledge is not available, we relax the assumption to approximate a process model. Instead, the RL-based controller with policy gradient search (PGS) that estimates the distribution of input–output relationship from historical output data of a system is introduced in this subsection.

One typical method to solve the process control problem without domain knowledge refers to PGS, which is based on the assumption that the system's output follows a distribution with control actions. We call this type of method RL-based control with PGS in process control problems. Suppose that system output $y_t$ follows a distribution with probability $p(y_t|u_t, y_{t-1}, \boldsymbol{\gamma_t})$, where $\boldsymbol{\gamma_t}$ denotes other parameters in the distribution that need to be estimated. Based on the total cost in Equation (2.1), the RL-based controller with PGS aims to minimize the expectation of total cost:

$$\min_{\boldsymbol{u}} E_{p(\boldsymbol{y}|\boldsymbol{u})}[\textstyle\sum_{t=1}^{T}(y_t - y^*)^2]. \tag{2.6}$$

The gradient descent method is used to minimize the cost in each period. We re-define the cost at period $t$ as $J_t(u_t) = E_{p(y_t;u_t)}C_t(y_t)$, where $C_t(y) = (y_t - y^*)^2$. To find the gradient of total cost over control actions ($J_{t\nabla}(u)$), the log-likelihood is used to simplify the solution. Thus, the gradient is calculated as:

$$J_{t\nabla}(u_t) = E_{p(y_t;u_t)}[C_t(y_t)\nabla_{u_t} \log(p(y_t;u_t))]. \tag{2.7}$$

Therefore, we propose Algorithm 2 to find the gradient descent direction $J_\nabla(u)$ and obtain the optimal solution by iterating control actions.

According to Algorithm 2, the RL-based controller with PGS searches a gradient descent direction of the control cost. The optimal cost value depends on the distribution of output at each period $t$, i.e., $p(y_t; u_t)$, which needs to be estimated from historical offline data. In general, uniform and normal distributions are accepted to describe the system outputs (Recht, 2019). For example, if the system output follows the normal distribution, we can formulate the probability density function as:

$$p(y_t; u_t) = \frac{1}{\sqrt{2\pi}\sigma} \exp\left(-\frac{(y_t - \beta(u_t - u_{t-1}))^2}{2\sigma^2}\right), \tag{2.8}$$

where $\sigma$ and $\beta$ are parameters that need to be estimated from the offline data. After parameter estimation, the gradient of control actions is calculated by Equation (2.7), and the iterations of control actions are presented by Algorithm 2.

In real applications, this policy gradient algorithm can deal with different kinds of disturbance processes, even though the disturbance variation is large. Since various distributions can be assumed in Algorithm 2 based on real data, we propose different simulation studies in Section 3 to verify the performance of the RL-based controller and compare it with traditional controllers. The nonlinear cases are elaborated in Section 4 to further show the advantages of RL-based controllers.

---

Algorithm 2. Online PGS algorithm in process control

---

Input: distribution $p(\cdot)$, parameters $T$, $y^*$, $\eta$, $\alpha$, $\boldsymbol{\gamma}$, and offline data $\boldsymbol{D}$

Estimate parameters $\boldsymbol{\gamma}$ in the distribution function $p(\cdot)$ according to data in $\boldsymbol{D}$.

Initialize $y_0$, $T$, $u_1^0$

**for** $t = 1:T$ **do**

    **repeat:**

$$g_t = C_t(y_t)\nabla_{u_t}\log(p(y_t;u_t))$$

$$u_t^{k+1} = u_t^k - \alpha g_t$$

    **Until** $\|u_t^{k+1} - u_t^k\| < \eta$

    Record the final convergent iteration action $u_t^*$ and $y_t^*$.

    $u_{t+1}^0 \leftarrow u_t^*$ if $t \neq T$

**end for**

Add the online control data in a run to $\boldsymbol{D}$

---

## 3. Classical linear simulation cases

To compare with the traditional controllers that are usually applied in linear process models, we first simulate two linear cases in the semiconductor manufacturing process. Section 3.1 mainly shows the performance of RL-based controllers with domain knowledge in the chemical mechanical planarization process. Section 3.2 illustrates the performance of RL-based controllers without domain knowledge in the deep reactive ion etching process.

### 3.1 *Application in the chemical mechanical planarization process*

Chemical mechanical planarization (CMP) is a crucial process in the semiconductor manufacturing process. Virtual metrology systems are often applied in the CMP process to remove the non-planar parts of the films. To simulate the CMP process, we first accept the simulation model in Ning et al. (1996)

and Chang et al. (2006), which is defined as:

$$y_t = A + Bu_t + \delta t + w_t, \tag{3.1}$$

where $y_t \in \mathbb{R}^{2\times 1}$ is the output vector that represents the removal rate and non-uniformity, respectively. $u_t \in \mathbb{R}^{4\times 1}$ is the control vector, that denotes the platen speed, back pressure, polish head downforce, and profile, respectively. Since Ning et al. (1996) proposed the EWMA controller for the CMP process, we use the same parameters for a fair comparison. The parameter matrices in Equation (3.1) are also assigned as $A = \begin{bmatrix} -138.21 \\ -627.32 \end{bmatrix}$, $B = \begin{bmatrix} 5.018 & -0.665 & 16.34 & 0.845 \\ 13.67 & 19.95 & 27.52 & 5.25 \end{bmatrix}$, and $\delta = \begin{bmatrix} -17 \\ -1.5 \end{bmatrix}$. The white noise $w_t$ is normally distributed with zero mean and covariance matrix $\Lambda = \begin{bmatrix} 665.64 & 0 \\ 0 & 5.29 \end{bmatrix}$. The total number of periods $T = 30$ (i.e., $t \in \{1, 2 \ldots 30\}$) represents the length of the CMP process. The objective is to keep $y_t$ at each time $t$ close to the target value of output $y^* = [1700, 150]^T$. The objective function to measure the control performance is to minimize the total cost calculated by the sum of square errors from the first period to the last period within a run, which is defined as:

$$\sum_{t=1}^{T}(y_t - y^*)^T(y_t - y^*). \tag{3.2}$$

To verify the performance of the RL-based controller, we make comparisons for RL-based controllers with traditional controllers. Moreover, to verify the advantages of the RL-based controller, it is also compared with the method of control optimization after parameter estimation.

## A. Comparison with the traditional controller

Since the EWMA controller is widely used in process control, we take the EWMA controller as a benchmark controller in the CMP process. To make the EWMA controller perform well, we consider the condition that all parameters in the process model in Equation (3.1) are known as the real values. While in the RL-based controller, only initial values need to be first given. Without loss of generality, we set the initial output as $y_0 = y^* = [1700, 150]^T$. In this case, due to the time-independent process model, the system outputs and control actions in different periods are used to estimate parameters in the process model. Thus, based on $N$ simulated sample paths (i.e., $N \times T$ samples in total), we estimate parameters $A$, $B$, and $\delta$, and make optimal control decision according to the estimated value of parameters $\hat{A}$, $\hat{B}$ and $\hat{\delta}$.

To comprehensively show the performance of two controllers, we simulate 1000 replications, and $N = 30$ sample paths are generated from Equation (3.1) in each replication. Figure 2 illustrates the distribution of the total cost under (a) the RL-based controller and (b) the EWMA controller in 500 replications, which is illustrated by boxplots. Since the cost variation of the first sample is much larger than other samples in RL-based controllers, we provide two figures to illustrate. The first figure in Figure 2 (a) shows the entire plot and the second shows the details from the second sample to the last sample. The *x*-axis shows the learning sample index, and the *y*-axis shows the total cost. Notably, the RL-based controller iterates from the first sample to the last sample according to Algorithm 1. By connecting the median of different samples by the black line, we find that the EWMA controller has a stable total cost in Figure 2(b) because parameters in the process model are exactly known as the truth. However, since the performance of the EWMA controller is imperfect to deal with linear drift, outliers exist in the EWMA controller.

For the RL-based controller, Figure 2(a) shows that the total cost reduces to a lower level than the EWMA controller at first and then fluctuates at the lower level with iterations of 30 sample paths. As the parameters are unknown in the process model of the RL-based controller, we estimated the parameters via Algorithm 1, thereby leading to performance fluctuation. However, the performance fluctuation is limited, and the outliers are less than the traditional EWMA controller.

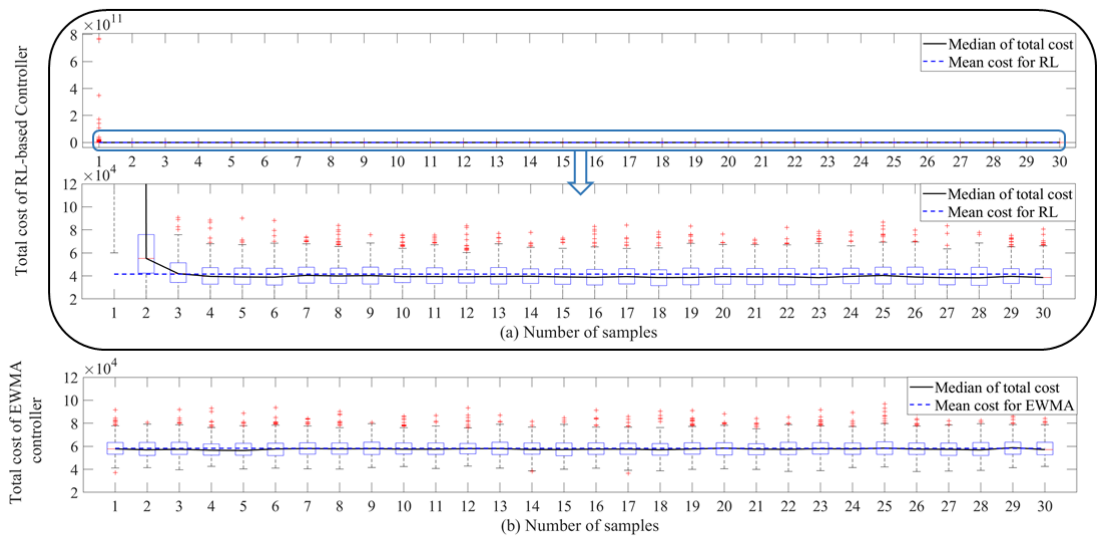

Figure 2. Boxplots of total cost under two controllers in 30 sample paths.

## B. Comparison with the control optimization after parameter estimation method

As shown in Proposition 1, the RL-based controller has a crucial advantage over the method of optimizing after parameter estimation (OAPE). In this part, we make a comparison of these two control methods based on the CMP process. Specifically, the mean square error (MSE) of the $N$th sample-path $MSE = \frac{1}{T}\sum_{t=1}^{T}(\boldsymbol{y_t} - \boldsymbol{y}^*)^T(\boldsymbol{y_t} - \boldsymbol{y}^*)$ is used to evaluate the control performance according to $N$ numbers of sample paths. To describe the robustness of the performance for these two control methods, Table 1 presents the mean and standard deviation of MSE according to 50 replications.

Table 1. Performance comparison of two control methods

| #of sample paths ($N$) | Mean of MSE | | Std. dev. of MSE | |
| --- | --- | --- | --- | --- |
| | RL-based control | OAPE control | RL-based control | OAPE control |
| 10 | 681.30 | 4228.74 | 173.51 | 6008.70 |
| 30 | 671.15 | 1670.92 | 143.82 | 1973.54 |
| 50 | 672.61 | 1019.64 | 182.87 | 552.91 |
| 100 | 673.02 | 904.67 | 163.08 | 463.24 |

As shown in Table 1 under the different number of sample paths, we find that the MSE and standard deviation of MSE of the RL-based controller are less than it the OAPE method. Particularly, the RL-based controller has better performance when the number of sample paths is limited. Furthermore, Figure 3 provides the boxplot of the total control cost defined in Equation (3.2) by using the RL-based controller and the OAPE method according to 50 replications, which shows the advantages of the RL-based controller intuitively.

## 3.2 Application in the deep reactive ion etching process

The deep reactive ion etching (DRIE) process is another important manufacturing process of semiconductor manufacturing. In existing research, the DRIE process is widely studied and used to compare the performance of different controllers. Due to the complex auto-correlation that exists in the disturbance process, the domain knowledge on proper correlation assumption is very limited, which hinders the use of the RL-based controller in Algorithm 1. Therefore, we propose the RL-based controllers without domain knowledge for process control in the DRIE process.

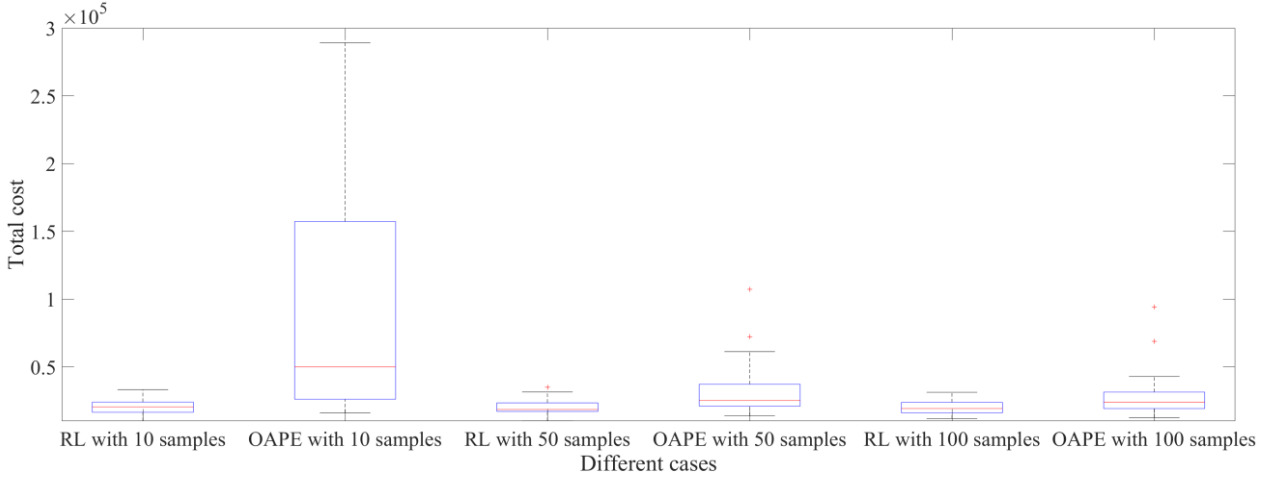

Figure 3. Comparison of RL-based controller and the OAPE method

To conduct a reasonable comparison with traditional controllers that need a proper process model with different disturbances (such as IMA, ARMA, and ARIMA processes) to describe the DRIE process, we refer to their process formulation to simulate data. Notably, the RL-based controller makes decisions according to the data merely. The widely accepted process model in the literature and also used in our simulation is as follows:

$$Y_t = a + bu_t + d_t, \qquad (3.3)$$

where $d_t$ is the disturbance at period $t$. Since IMA and ARMA processes can be treated as special cases of the ARIMA process, we focus on the ARIMA disturbance process directly. i.e., we have the formulation of $d_t$ as follows:

$$\begin{cases} d_t = d_{t-1} + \Delta d_t \\ \Delta d_t = \phi \Delta d_{t-1} + w_t - \theta w_{t-1}, \end{cases} \qquad (3.4)$$

where $\phi, \theta \in (0,1)$, and $w_t$ is white noise.

Existing works claimed that the traditional EWMA controller cannot deal with ARIMA disturbance very well and proposed modified controllers, such as VEWMA controller (Tseng et al., 2003) and GHR controller (He et al., 2009). He et al. (2009) numerically showed that the GHR controller had better control performance than the VEWMA controller and EWMA controller. Hence, we use the GHR controller as the quasi-optimal controller to compare with our RL-based controllers. To make a fair comparison, we use the same parameters in Equations (3.3) and (3.4) as He et al. (2009)

to generate data, i.e., $a = 91.7$, $b = -1.8$, $\theta = 0.5$, and $\varphi = 0.6$. Moreover, the total number of periods is $T = 80$, and the target value of $y^* = 90$.

Since domain knowledge is not available, we use the PGS method to make control decisions according to the distribution of system output, which is elaborated in Algorithm 2, where the historical offline data is necessary for the distribution estimation. Due to the unknown process model, we try to approximate the output only based on the distribution of offline sample paths. In this case, the generalized Brownian Motion is used to approximate. Then, the online distribution approximation of the system output $y_t$ follows a normal distribution with mean value as: $\mu_{y_t|y_{t-1}} = y_{t-1} + \beta_t(u_t - u_{t-1})$, and variance $\text{var}_{y_t|y_{t-1}} = v(t)$. We assume that control actions can only influence the mean value of system outputs, and variance only depends on the operation time of the manufacturing system (constant variance can also be accepted based on real data). In our study, a time-dependent variance function $\text{var}_{y_t|y_{t-1}} = \gamma^2 t$ is accepted. Moreover, parameters $\beta$ and $\gamma$ are estimated from the offline data.

According to Algorithm 2, $\hat{\beta}$ and $\hat{\gamma}$ are first estimated by 1000 offline sample paths. Then the control actions are iterated to reduce the control cost. As a result, we obtain the system output of the RL-based controller and compare it with the GHR controller, which is shown in Figure 4:

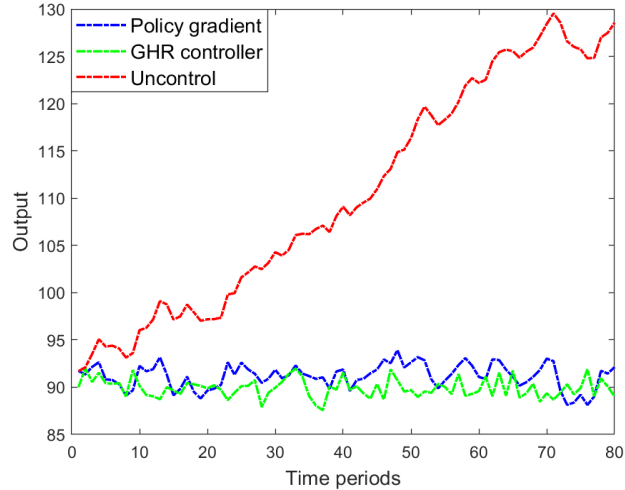

Figure 4. Results of three control methods based on ARIMA disturbance.

In Figure 4, the process model and parameters are exactly known for the GHR controller, while the RL-based controller with PGS relaxes the process model assumptions, which only estimates the output distribution from offline data. As a result, the performances of the GHR controller and RL-based controller with PGS are comparable in the ARIMA disturbance process. To show the comprehensive performance of both controllers, 30 replications are simulated in the boxplot of total costs. Figure 5 displays the boxplots of the total cost of two controllers:

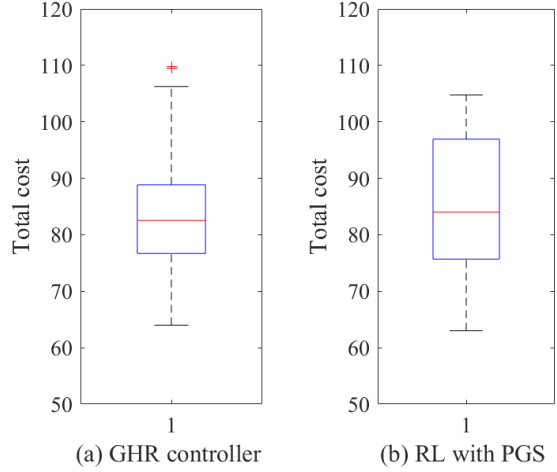

Figure 5. Distributions of the total cost based on GHR controller and RL-based controller with PGS.

As shown in Figure 5, the GHR and RL-based controller with PGS have comparable performance. However, the RL-based controller only depends on historical offline data and relaxes the process model assumption, while the GHR controller relies on the explicit process model in Equations (3.3) and (3.4). The superiority of the RL-based controller is obvious if the process model is unavailable.

To recognize the rationality of the RL-based controller, we make a theoretical analysis of the difference between the approximate process model with the underlying ground truth model in this case. For the variance of these two stochastic processes, according to the linear process model with an ARIMA disturbance process, we have the variance of system output as follows:

$$\text{var}_{y_t} = \left( \sum_{i=1}^{t-1}(t-i)\left(\phi^{i-1}(\phi-\theta)\right)^2 + t \right)\sigma^2, \qquad (3.5)$$

where $\sigma^2$ is the variance of white noise $w_t$ in Equation (3.4). The variance difference of system output $\Delta\text{var}_{y_t}$ with ARIMA disturbances and traditionally Brownian motion process is $\Delta\text{var}_{y_t} = (\phi-\theta)^2 S_t \sigma^2$, where $S_t = \sum_{i=1}^{t-1}(t-i)\left(\phi^{i-1}\right)^2$. Since $S_t$ increases with the period $t$, and so does $\Delta\text{var}_{y_t}$. Therefore, the time-dependent variance function $v(t) = \text{var}_{y_t|y_{t-1}} = \gamma^2 t$ is accepted and more accurate than the traditional Brownian motion process (i.e. variance function $v(t)$ is a constant function) to describe the manufacturing process with ARIMA disturbances.

Besides the first-order differential sequence of the ARIMA process, the RL-based controller with PGS can solve higher-order differential sequences. For example, if $d_t$ follows ARIMA (1,2,1), we can use the stochastic process $\tilde{d}_t \sim N(0, \sigma^2 \sum_{j=1}^{t} j)$ to approximate the second-order ARIMA disturbance.

One of the most potential advantages of RL-based controllers with PGS is they are data-driven controllers that can handle more complicated process models. Therefore, besides traditional simulation cases, we proposed other complicated simulation studies, to further verify the performance of RL-based controllers.

## 4. Other nonlinear simulation cases

Besides the widely used linear process models, nonlinear models are proposed to describe the semiconductor manufacturing processes. In this section, we focus on the evaluation of RL-based controllers for different nonlinear models. In subsection 4.1, domain knowledge is available for a nonlinear approximate model. While in 4.2, we introduce cases with complicated nonlinear process models, which are hard to be approximated by domain knowledge.

### 4.1 *Application in a nonlinear CMP process*

According to the experiment tool presented by Khuri (1996), Del Castillo and Yeh (1998) simulated the CMP process by a nonlinear process model. In this case, three control variables are considered: back pressure downforce ($u_1$), platen speed ($u_2$), and the slurry concentration ($u_3$). The system outputs to reflect the manufacturing quality are removal rate ($y_1$) and within-wafer standard deviation ($y_2$). Similar to Section 3.1, control decision aims to adjust the system output in each period close to the target level, which is defined as $y_1^* = 2200$ and $y_2^* = 400$. We use the same parameters as Del Castillo and Yeh (1998), which are estimated from the results of a 32-wafer experimental design. The process model is:

$$y_1 = 2756.5 + 547.6u_1 + 616.3u_2 - 126.7u_3 - 1109.5u_1^2 - 286.1u_2^2 + 989.1u_3^2 - 52.9u_1u_2 \\ - 156.9u_1u_3 - 550.3u_2u_3 - 10t + \varepsilon_{1t}$$

$$y_2 = 746.3 + 62.3u_1 + 128.6u_2 - 152.1u_3 - 289.7u_1^2 - 32.1u_2^2 + 237.7u_3^2 - 28.9u_1u_2 \\ - 122.1u_1u_3 - 140.6u_2u_3 + 1.5t + \varepsilon_{2t},$$

(4.1)

where $\varepsilon_{1t} \sim N(0, 60^2)$ and $\varepsilon_{2t} \sim N(0, 30^2)$ are the random errors.

If a quadratic process model can be guided from domain knowledge, we use Algorithm 1 to estimate the parameters in the model and optimize the control action alternately. Figure 6 illustrates the result of controlled system outputs within a run from time 0 to $T$ via Algorithm 1 after iterations of 1000 sample paths, where the parameters in the approximate process and optimized control actions are both convergent.

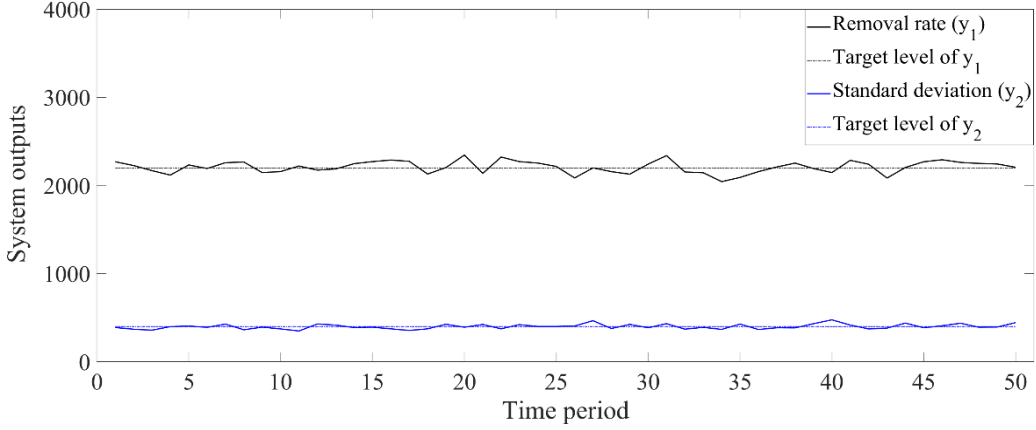

Figure 6. System output of RL-based controller for quadratic process model

To evaluate the performance of RL-based controllers, we define the error ratio $\rho$ of system errors with the target level as $\rho = \frac{y_t - y^*}{y^*}$. Figure 7 presents the error ratio $\rho$ by boxplot according to 50 sample paths. We find most error ratios are less than 0.1 for $y_1$ and less than 0.2 for $y_2$, since $y_2$ has a relatively larger variation given $\varepsilon_{2t}$ with respect to the value of $y_1$ than $\varepsilon_{1t}$. In summary, as long as the domain knowledge can guide a reliable approximate model, which is not restricted to linear or polynomial models, the RL-based controller in Algorithm 1 can handle the control problem by estimating parameters and optimizing control actions.

## 4.2 *Process model approximated by stochastic processes*

When the process model is unavailable and hard to be approximated by domain knowledge, finding a proper approximate model is challenging. RL-based controller with PGS in Algorithm 2 can analyze the relationship between control actions and system output from historical offline data and find the optimal control actions for unknown complicated nonlinear models.

In existing simulation studies for the semiconductor manufacturing process, besides linear and quadratic models, few complicated process model is proposed in the simulation cases. However, in our work, to further prove the advantages of RL-based controllers, we introduce two stochastic processes to simulate the manufacturing process data for RL-based controllers.

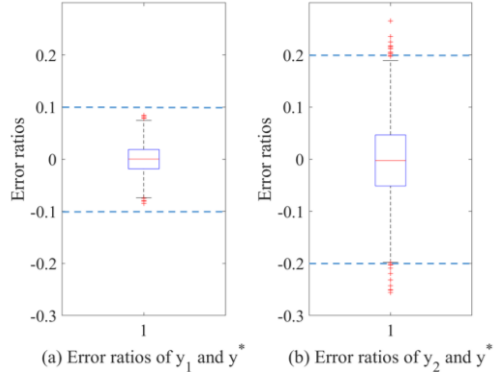

Figure 7. Boxplot of error ratios $\rho$ between system errors and $y^*$

## A. Wiener process

It is widely accepted that the manufacturing system has a random drift if there is no control action. For the validation purpose, we simulate the system output data by the Wiener process, which is a widely discussed stochastic process. According to Peng and Tseng (2009), we simulate the process data by the expression as follows:

$$y_t = y_0 + vt + \sigma B(t), \qquad (4.2)$$

where $y_0$ is the initial system output, $v$ and $\sigma$ are the drift and diffusion parameters, respectively.

Similar to Section 3.2, we use Algorithm 2 to optimize the control actions. 30 replications are simulated, and Figure 8 illustrates the error ratio of the system output with and without control. It is obvious that the RL-based controller reduces the error and keeps the system output to the target level ($y^* = 90$).

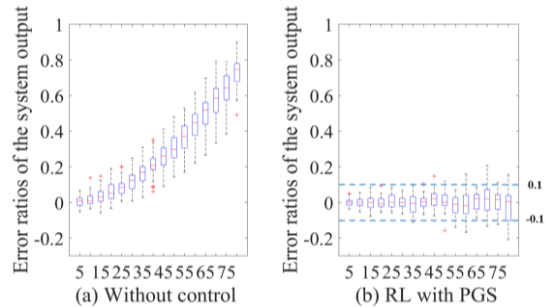

Figure 8. System output with the Wiener-process disturbance

## B. Gamma process

Another well-known stochastic process is the Gamma process. Therefore, we also use the Gamma process to verify the performance of the RL-based controller with PGS. According to the formulation in Cheng et al. (2018), we have the process model for data simulation as $y_t = y_{t-1} + \Delta y_t$, where $\Delta y_t$ follows a gamma distribution with probability density function:

$$f_{\Delta y}(y) = \frac{\beta^\alpha y^{\alpha-1}}{\Gamma(\alpha)} e^{-\beta y}, \qquad (4.3)$$

where $\alpha$ is shape parameter, $\beta$ is the scale parameter and $\Gamma(\cdot)$ is Gamma function. We use the same parameters in their numerical study from Cheng et al. (2018), i.e. $\alpha = 0.36$, and $\beta = 0.64$. Without loss of generality, the total number of manufacturing periods $T = 80$, and the target output $y^* = 90$. Figure 9 illustrates the error ratio of the system output based on gamma process models. Similar to the Wiener process, we find that RL-based controllers are also efficient in terms of the gamma process. Almost all absolute values of error ratios are less than 0.1.

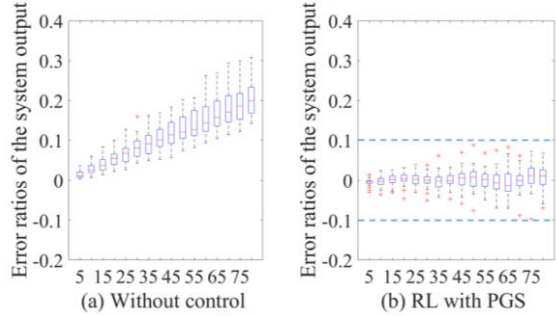

Figure 9. Error ratios of system output with the Gamma-process disturbance

Table 2 summarizes the mean and standard deviation of MSE for these two stochastic simulation processes by 30 replications. To show the advantages of RL-based controllers intuitively, we calculate the mean and standard deviation of MSE, and also the ratios (shown in the bracket) of two control methods (i.e. without control and RL-based controller). We find that the RL-based controller can reduce more than 95% of the process cost comparing with the scenario without control. Meanwhile, the standard deviations of control cost are also reduced to 5%-8% by the RL-based controller.

Table 2. Control cost in different simulation cases

| Simulation cases | Control cases | Mean of MSE | Std. dev. of MSE |
|---|---|---|---|
| (1) Wiener process model | Without control | 992.22 | 331.34 |
| | RL-based controller | 21.86 | 16.86 |
| | | (0.022) | (0.051) |
| (2) Gamma process model | Without control | 125.80 | 61.93 |
| | RL-based controller | 4.37 | 4.63 |
| | | (0.035) | (0.075) |

## 4.3 *Discussion*

After two parts of simulation studies, we conclude that: (1) according to the general disturbance setting in existing literature, RL-based controllers can solve the control problem without process model assumptions, and outperform or at least have comparable performance with traditional controllers. (2) To further verify RL-based controllers' superiority, we use other process models besides the linear model to describe the input–output relationship such as quadratic models. In addition, we refer to two stochastic process models (Wiener process and Gamma process) to further validate the proposed RL-based controller with PGS.

If the domain knowledge is available for approximation of unknown process models (not restricted to linear model), the RL-based controller in Algorithm 1 outperforms the traditional EWMA controller to deal with the system drift. Even if the parameters in the approximate process model are unknown, the total cost of the RL-based controller decreases within limited samples.

If the process model is difficult to approximate, the RL-based controller in Algorithm 2 can still deal with various disturbances. Based on the ARIMA process, RL-based controllers even have comparable performance with GHR controllers, where exactly true process models and parameters are used. Moreover, other stochastic process models are also analyzed. Since traditional controllers are invalid and hard to be compared in this setting, we evaluate the control results of RL-based controllers by using error ratios between system output and the target level. The results show that most error ratios of the RL-based controller are less than 0.1.

# 5. Conclusions

Process control is an important problem in the semiconductor manufacturing process to reduce process variation. Considering the presence of various disturbances in the manufacturing environments, our work aims to obtain the optimal control action according to the historical offline data and real-time output of the system. Different from traditional control methods, which focus on building a linear process model to describe the input–output relationship, we propose RL-based controllers and relax the assumption of the process model. Compared with traditional controllers, RL-based controllers are suitable for more complicated applications, and not restricted to specific process models.

Based on the availability of domain knowledge to approximated a process model, we propose two different RL-based control algorithms and discuss some theoretical properties based on the widely accepted linear process models. Two sections of simulations prove that RL-based controllers are not only better or at least comparable with traditional controllers in linear simulation cases, but also show great potential to deal with more complicated cases. In the future, improvements of RL-based controllers can be further made, such as variations reduction of system outputs under model-free cases.

# Appendix

## *Appendix A. Proof of Theorem 1*

Based on Assumption 2.1, we have the approximate process model as $\hat{f}_t(\cdot|\boldsymbol{\Theta}_t) = \boldsymbol{\Theta}_t\boldsymbol{\Psi}(\boldsymbol{\tau}_t) + e_t$, where $\boldsymbol{\tau}_t$ is the control and disturbance trajectory is observed from the offline data. If $\boldsymbol{\Psi}(\cdot)$ is available according to domain knowledge, the parameters $\boldsymbol{\Theta}_t$ in $\hat{f}_t(\cdot|\boldsymbol{\Theta}_t)$ can be estimated by the least square method. As a result, we have the parameter estimator of $\boldsymbol{\Theta}_t$, i.e., $\widehat{\boldsymbol{\Theta}}_t = (X_t^T X_t)^{-1} X_t^T y_t$. The estimation error is given by $\widehat{\boldsymbol{\Theta}}_t - \boldsymbol{\Theta}_t = (X_t^T X_t)^{-1} X_t^T e_t$, which is directly proportional to $1/N$, and the total number of samples in period $t$ is $N$.

## *Appendix B. Proof of Theorem 2*

Assumption 2.2 proposes the linear model for $\hat{f}_t(\cdot)$, we have $y_t = b_t u_t + \boldsymbol{Y}_t K_t\left(\boldsymbol{\tau}_t^{[\backslash u_t]}\right) + e_t$, where $e_t$ is the random Gaussian error defined as $e_t \sim N(0, \sigma^2)$. We denote that the set of parameters to be estimated as $\boldsymbol{\Theta}_t = [b_t, \boldsymbol{\Upsilon}_t^T]^T$. Then we reformulate the approximate model as $y_t = b_t u_t + c_t + e_t$, where $c_t = \boldsymbol{\Upsilon}_t \boldsymbol{K}_t\left(\boldsymbol{\tau}_t^{[\backslash u_t]}\right)$.

At each period $t$, the online control decision is made according to the parameter estimators $\hat{b}_t$ and $\hat{\boldsymbol{\Upsilon}}_t$. We have the optimal control policy $u_t^* = \frac{y^* - \hat{c}_t}{\hat{b}_t}$, and $\hat{c}_t = \hat{\boldsymbol{\Upsilon}}_t \boldsymbol{K}_t\left(\boldsymbol{\tau}_t^{[\backslash u_t]}\right)$, where $\boldsymbol{\tau}_t^{[\backslash u_t]} = [u_{t-1} \ldots u_1, d_{t-1} \ldots d_1, 1]^T$ and $\boldsymbol{K}_t(\cdot)$ are known. Based on the maximum likelihood estimation, we can conclude that the parameter estimators follow the multi-normal distribution as: $\hat{\boldsymbol{\Theta}}_t = \begin{bmatrix} \hat{b}_t \\ \hat{\boldsymbol{\Upsilon}}_t \end{bmatrix} \sim MN\left(\begin{bmatrix} b_t \\ \boldsymbol{\Upsilon}_t \end{bmatrix}, \boldsymbol{\Sigma}_\Theta\right)$, and the covariance matrix is formulated as:

$$\boldsymbol{\Sigma}_\Theta = \sigma^2 [X_t^T X_t]^{-1}$$
$$= \sigma^2 \begin{bmatrix} \sum_n u_{t_n}^2 & \sum_n \left[u_{t_n} \boldsymbol{K}_t\left(\boldsymbol{\tau}_t^{[\backslash u_t]}\right)\right] \\ \sum_n \left[\left(\boldsymbol{K}_t\left(\boldsymbol{\tau}_t^{[\backslash u_t]}\right)\right)^T u_{t_n}\right] & \sum_n \left[\left(\boldsymbol{K}_t\left(\boldsymbol{\tau}_t^{[\backslash u_t]}\right)\right)^T \boldsymbol{K}_t\left(\boldsymbol{\tau}_t^{[\backslash u_t]}\right)\right] \end{bmatrix}^{-1}, \quad (B.1)$$

where $X_t = \begin{bmatrix} u_{t_1} & \boldsymbol{K}_t\left(\boldsymbol{\tau}_{t_1}^{[\backslash u_t]}\right) \\ u_{t_2} & \boldsymbol{K}_t\left(\boldsymbol{\tau}_{t_2}^{[\backslash u_t]}\right) \\ \ldots & \ldots \\ u_{t_n} & \boldsymbol{K}_t\left(\boldsymbol{\tau}_{t_N}^{[\backslash u_t]}\right) \end{bmatrix} = [\boldsymbol{U}_t \quad \boldsymbol{K}_t]$, and $n \in \{1, 2, \ldots N\}$ is the index of the sample path. And $\boldsymbol{U}_t \in \mathbb{R}^{N \times 1}$ denotes the control actions in historical sample paths, $\boldsymbol{K}_t \in \mathbb{R}^{N \times m}$ denotes the function data of historical trajectories of the offline sample paths, $N$ is the total number of sample paths for parameter estimation, and $m$ is the dimension of function $\boldsymbol{K}_t(\cdot)$. To further simplify the notations, we reformulate Equation (B.1) as matrix form:

$$\boldsymbol{\Sigma}_\Theta = \sigma^2 \begin{bmatrix} \boldsymbol{U}_t^T \boldsymbol{U}_t & \boldsymbol{U}_t^T \boldsymbol{K}_t \\ \boldsymbol{K}_t^T \boldsymbol{U}_t & \boldsymbol{K}_t^T \boldsymbol{K}_t \end{bmatrix}^{-1}.$$

According to Khan (2008), we have the expression of $\boldsymbol{\Sigma}_\Theta$ as

$$\boldsymbol{\Sigma}_\Theta = \sigma^2 \begin{bmatrix} \dfrac{1}{\boldsymbol{U}_t^T \boldsymbol{U}_t - \boldsymbol{U}_t^T \boldsymbol{K}_t (\boldsymbol{K}_t^T \boldsymbol{K}_t)^{-1} \boldsymbol{K}_t^T \boldsymbol{U}_t} & \dfrac{-\boldsymbol{U}_t^T \boldsymbol{K}_t \left((\boldsymbol{K}_t^T \boldsymbol{K}_t)^{-1}\right)^T}{\boldsymbol{U}_t^T \boldsymbol{U}_t - \boldsymbol{U}_t^T \boldsymbol{K}_t (\boldsymbol{K}_t^T \boldsymbol{K}_t)^{-1} \boldsymbol{K}_t^T \boldsymbol{U}_t} \\ \dfrac{-(\boldsymbol{K}_t^T \boldsymbol{K}_t)^{-1} \boldsymbol{K}_t^T \boldsymbol{U}_t}{\boldsymbol{U}_t^T \boldsymbol{U}_t - \boldsymbol{U}_t^T \boldsymbol{K}_t (\boldsymbol{K}_t^T \boldsymbol{K}_t)^{-1} \boldsymbol{K}_t^T \boldsymbol{U}_t} & (\boldsymbol{K}_t^T \boldsymbol{K}_t)^{-1} + \dfrac{(\boldsymbol{K}_t^T \boldsymbol{K}_t)^{-1} \boldsymbol{K}_t^T \boldsymbol{U}_t \boldsymbol{U}_t^T \boldsymbol{K}_t ((\boldsymbol{K}_t^T \boldsymbol{K}_t)^{-1})^T}{\boldsymbol{U}_t^T \boldsymbol{U}_t - \boldsymbol{U}_t^T \boldsymbol{K}_t (\boldsymbol{K}_t^T \boldsymbol{K}_t)^{-1} \boldsymbol{K}_t^T \boldsymbol{U}_t} \end{bmatrix}$$
(B.2)

During the online control process, we have $\hat{c}_t = \hat{\boldsymbol{\Upsilon}}_t \boldsymbol{K}_t\left(\boldsymbol{\tau}_t^{[\backslash u_t]}\right)$. Since the online historical trajectory $K_t\left(\tau_t^{[\backslash u_t]}\right)$ and $y^*$ are already known, we can easily conclude that $y^* - \hat{c}_t$ and $\hat{b}_t$ follow the multivariate normal distribution as $\begin{bmatrix} y^* - \hat{c}_t \\ \hat{b}_t \end{bmatrix} \sim MN\left(\begin{bmatrix} y^* - c_t \\ b_t \end{bmatrix}, \Sigma\right)$, and $\Sigma$ is expressed as:

$$\Sigma = \sigma^2 \begin{bmatrix} K_t\left(\tau_t^{[\backslash u_t]}\right)\left((K_t^T K_t)^{-1} + \dfrac{(K_t^T K_t)^{-1} K_t^T U_t U_t^T K_t \left((K_t^T K_t)^{-1}\right)^T}{U_t^T U_t - U_t^T K_t (K_t^T K_t)^{-1} K_t^T U_t}\right)\left(K_t\left(\tau_t^{[\backslash u_t]}\right)\right)^T & \dfrac{K_t\left(\tau_t^{[\backslash u_t]}\right)(K_t^T K_t)^{-1} K_t^T U_t}{U_t^T U_t - U_t^T K_t (K_t^T K_t)^{-1} K_t^T U_t} \\ \dfrac{U_t^T K_t\left((K_t^T K_t)^{-1}\right)^T K_t\left(\tau_t^{[\backslash u_t]}\right)^T}{U_t^T U_t - U_t^T K_t (K_t^T K_t)^{-1} K_t^T U_t} & \dfrac{1}{U_t^T U_t - U_t^T K_t (K_t^T K_t)^{-1} K_t^T U_t} \end{bmatrix}$$

(B.3)

To simplify the notation, we use $\mu_1$ and $\mu_2$, $\sigma_1$ and $\sigma_2$ to denote the mean values and standard deviations of $y^* - \hat{c}_t$ and $\hat{b}_t$ respectively, $\sigma_{12}$ denotes the covariance of $y^* - \hat{c}_t$ and $\hat{b}_t$, and their correlation coefficient is denoted as $\rho$.

According to the conclusion of Hinkley (1969), without loss of generality, assuming that coefficient $b_t > 0$, we have the probability distribution function of $u_t = \dfrac{y^* - \hat{c}_t}{\hat{b}_t}$ based on the distribution of $y^* - \hat{c}_t$ and $\hat{b}_t$ (if $b_t < 0$, we get the distribution of $-u_t$):

$$f(u) = \dfrac{\beta(u)\gamma(u)}{\sqrt{2\pi}\sigma_1\sigma_2\alpha^3(u)}\left[\Phi\left\{\dfrac{\beta(u)}{\sqrt{1-\rho^2}\alpha(u)}\right\} - \Phi\left\{-\dfrac{\beta(u)}{\sqrt{1-\rho^2}\alpha(u)}\right\}\right] + \dfrac{\sqrt{1-\rho^2}}{\pi\sigma_1\sigma_2\alpha^2(u)}\exp\left\{-\dfrac{c}{2(1-\rho^2)}\right\}, \quad (B.4)$$

where $\alpha(u) = \left(\dfrac{u^2}{\sigma_1^2} - \dfrac{2\rho u}{\sigma_1\sigma_2} + \dfrac{1}{\sigma_2^2}\right)^{\frac{1}{2}}$, $\beta(u) = \dfrac{\mu_1 u}{\sigma_1^2} - \dfrac{\rho(\mu_1+\mu_2 u)}{\sigma_1\sigma_2} + \dfrac{\mu_2}{\sigma_2^2}$, $c = \dfrac{\mu_1^2}{\sigma_1^2} - \dfrac{2\rho\mu_1\mu_2}{\sigma_1\sigma_2} + \dfrac{\mu_1^2}{\sigma_2^2}$, $\gamma(u) = \exp\left\{\dfrac{\beta^2(u) - c\alpha^2(u)}{2(1-\rho^2)\alpha^2(u)}\right\}$, and $\Phi\{\cdot\}$ denotes the cumulative distribution function of standard normal distribution. Furthermore, the cumulated distribution function of $u_t$ is:

$$F(u) = prob(u_t \leq u) = L\left\{\dfrac{\mu_1 - \mu_2 u}{\sigma_1\sigma_2\alpha(u)}, -\dfrac{\mu_2}{\sigma_2}; \dfrac{\sigma_2 u - \rho\sigma_1}{\sigma_1\sigma_2\alpha(u)}\right\} + L\left\{-\dfrac{\mu_1 - \mu_2 u}{\sigma_1\sigma_2\alpha(u)}, \dfrac{\mu_2}{\sigma_2}; \dfrac{\sigma_2 u - \rho\sigma_1}{\sigma_1\sigma_2\alpha(u)}\right\},$$

(B.5)

where $L(h, k; r) = \dfrac{1}{2\pi\sqrt{(1-r^2)}} \int_h^\infty \int_k^\infty \exp\left\{-\dfrac{x^2 - 2\gamma xy + y^2}{2(1-r^2)}\right\} dx\, dy$. Hinkley also provided an approximation normal distribution for $F(u)$, which is denoted as

$$F^*(u) = \Phi\left\{\dfrac{\mu_2 u - \mu_1}{\sigma_1\sigma_2\alpha(u)}\right\}. \quad (B.6)$$

Moreover, the bound of is also proposed: $|F(u) - F^*(u)| \leq \Phi\left(-\dfrac{\mu_2}{\sigma_2}\right)$. When $\dfrac{\mu_2}{\sigma_2} \to \infty$, $F(u)$ can be approximated by $F^*(u)$ well.

We first analyze $F^*(u)$. Suppose that we use $F^*(u)$ to approximate the real cumulative distribution function $F(u)$ and make a decision based only on $F^*(u)$. To distinguish the difference between $F(u)$ and $F^*(u)$, we use $\tilde{u}$ to denote the random variable in $F^*(\cdot)$. From Equation (B.6), it is obvious that we can construct a standard normal distribution random variable $Z = \frac{\mu_2 \tilde{u} - \mu_1}{\sigma_1 \sigma_2 \alpha(\tilde{u})} \sim N(0,1)$. According to Chebyshev inequality, we have $P(|Z - E(Z)| > \varepsilon) \leq \frac{var(Z)}{\varepsilon^2}$, i.e.

$$P\left(\left|\frac{\mu_2 \tilde{u} - \mu_1}{\sigma_1 \sigma_2 \alpha(\tilde{u})} - \frac{\mu_2 u^* - \mu_1}{\sigma_1 \sigma_2 \alpha(u^*)}\right| > \varepsilon\right) \leq \frac{1}{\varepsilon^2}. \tag{B.7}$$

Then we analyze the function $\sigma_1 \sigma_2 \alpha(\tilde{u})$. By substitute $\alpha(\tilde{u})$, we have $\sigma_1 \sigma_2 \alpha(\tilde{u}) = \text{std}\left(\hat{b}_t \tilde{u} - (y^* - \hat{c}_t)\right)$. To analyze the standard error of $\hat{b}_t \tilde{u} - (y^* - \hat{c}_t)$, we analyze the $\text{var}\left(\hat{b}_t \tilde{u} - (y^* - \hat{c}_t)\right)$ instead. Let $g(\tilde{u}) := \text{var}\left(\hat{b}_t \tilde{u} - (y^* - \hat{c}_t)\right)$, and we have:

$$\begin{aligned} g(u) &= E\left(\left(\hat{b}_t u - (y^* - \hat{c}_t)\right)^2\right) - E\left(\hat{b}_t u - (y^* - \hat{c}_t)\right)^2 = E\left(\hat{b}_t^2 u^2 + (y^* - \hat{c}_t)^2 - 2\hat{b}_t u(y^* - \hat{c}_t)\right) - \left(E(\hat{b}_t)u - E(y^* - \hat{c}_t)\right)^2 \\ &= u^2(b_t^2 + \sigma_2^2) - 2u\left(b_t(y^* - c_t) + cov(b_t, (y^* - c_t))\right) + (y^* - c_t)^2 + \sigma_1^2 - \left(u^2 b_t^2 + (y^* - c_t)^2 - 2b_t u(y^* - c_t)\right) \\ &= u^2 \sigma_2^2 - 2u \cdot \sigma_{12} + \sigma_1^2. \end{aligned}$$

(B.8)

It is easy to find that $g(\tilde{u})$ has a minimal value, i.e., when $\tilde{u} = \frac{\sigma_{12}}{\sigma_2^2} = K_t\left(\tau_t^{[\backslash u_t]}\right)\left(K_t^T K_t\right)^{-1} K_t^T U_t := \bar{u}_t$, we have $g_{min} = \frac{\sigma_2^2 \sigma_1^2 - \sigma_{12}^2}{\sigma_2^2}$, thereby indicating $g(\tilde{u}) \geq g(\bar{u}_t)$.

Therefore, we have $\left|\frac{\mu_2 \bar{u}_t - \mu_1}{\sqrt{g(\bar{u}_t)}} - \frac{\mu_2 u^* - \mu_1}{\sqrt{g(u^*)}}\right| = \left|\frac{\mu_2 \bar{u}_t - \mu_1}{\sqrt{g(\bar{u}_t)}} - \frac{\mu_2 u^* - \mu_1}{\sqrt{g(\bar{u})}}\right| = \frac{\mu_2 \sigma_2}{\sqrt{\sigma_2^2 \sigma_1^2 - \sigma_{12}^2}} |\bar{u}_t - u^*|$. The first equation is valid due to $\mu_2 u^* - \mu_1 = b_t u^* - (y^* - c_t) = 0$. According to Equation (B.7), we have

$$P\left(\frac{\mu_2 \sigma_2}{\sqrt{\sigma_2^2 \sigma_1^2 - \sigma_{12}^2}} |\bar{u}_t - u^*| > \varepsilon\right) \leq \frac{1}{\varepsilon^2}. \tag{B.9}$$

Let $\frac{\varepsilon \sqrt{\sigma_2^2 \sigma_1^2 - \sigma_{12}^2}}{\mu_2 \sigma_2} = \eta$, $\varepsilon = \frac{\mu_2 \sigma_2}{\sqrt{\sigma_2^2 \sigma_1^2 - \sigma_{12}^2}} \eta$, according to distribution $F^*(\cdot)$, we have

$$P(|\bar{u}_t - u^*| > \eta) < \frac{\sigma_2^2 \sigma_1^2 - \sigma_{12}^2}{(\mu_2 \sigma_2 \eta)^2}. \tag{B.10}$$

Then we focus on the real distribution function $F(u)$. Since the upper bound of difference between $F(u)$ and $F^*(u)$ is proved as $\Phi\left(-\frac{\mu_2}{\sigma_2}\right)$, we have

$$\begin{cases} F(u^* + \eta) \leq F^*(u^* + \eta) + \Phi\left(-\frac{\mu_2}{\sigma_2}\right) \\ F(u^* - \eta) \geq F^*(u^* - \eta) - \Phi\left(-\frac{\mu_2}{\sigma_2}\right). \end{cases} \tag{B.11}$$

According to Equation (B.11), it is obvious that:

$$P(|\bar{u}_t - u^*| > \eta) = 1 - P(|\bar{u}_t - u^*| \leq \eta) = 1 - P(u^* - \eta \leq \bar{u} \leq u^* + \eta)$$
$$= 1 - (F^*(u^* + \eta) - F^*(u^* - \eta)) < \frac{\sigma_2^2 \sigma_1^2 - \sigma_{12}^2}{(\mu_2 \sigma_2 \eta)^2},$$

and we have $F^*(u^* + \eta) - F^*(u^* - \eta) > 1 - \frac{\sigma_2^2 \sigma_1^2 - \sigma_{12}^2}{(\mu_2 \sigma_2 \eta)^2}$. Combining with Equation (B.11), we have $F(u^* + \eta) - F(u^* - \eta) \geq 1 - \frac{\sigma_2^2 \sigma_1^2 - \sigma_{12}^2}{(\mu_2 \sigma_2 \eta)^2} + 2\Phi\left(-\frac{\mu_2}{\sigma_2}\right)$. Therefore, we have $P(|\bar{u}_t - u^*| \leq \eta) \geq 1 - \left(\frac{\sigma_2^2 \sigma_1^2 - \sigma_{12}^2}{(\mu_2 \sigma_2 \eta)^2} + 2\Phi\left(-\frac{\mu_2}{\sigma_2}\right)\right)$, and $P(|\bar{u}_t - u^*| > \eta) \leq \left(\frac{\sigma_2^2 \sigma_1^2 - \sigma_{12}^2}{(\mu_2 \sigma_2 \eta)^2} + 2\Phi\left(-\frac{\mu_2}{\sigma_2}\right)\right)$. According to the linear relationship of input–output relationship, we have $E(y_t(\bar{u}_t)) = b_t \bar{u}_t + c_t$, and $E(y_t(u^*)) = b_t u^* + c_t = y_t^*$. It is also easy to prove that $P(|E(y_t(\bar{u}_t)) - y_t^*| > \eta) \leq \frac{\sigma_2^2 \sigma_1^2 - \sigma_{12}^2}{(\sigma_2 \eta)^2} + 2\Phi\left(-\frac{\mu_2}{\sigma_2}\right)$.